\documentclass{article}
\usepackage{ijcai23}
\usepackage{array}

\usepackage{times}
\usepackage{soul}
\usepackage{url}
\usepackage[hidelinks]{hyperref}
\usepackage[utf8]{inputenc}
\usepackage[small]{caption}
\usepackage{graphicx}
\usepackage{amsmath}
\usepackage{amsthm}
\usepackage{booktabs}
\usepackage{algorithm}
\usepackage{algorithmic}
\usepackage[switch]{lineno}

\usepackage{multirow}
\usepackage{times,graphicx}
\usepackage[edges]{forest}
\usepackage{xcolor} 
\definecolor{customgreen}{HTML}{70AD47} 
\definecolor{customorange}{HTML}{E6990E} 
\definecolor{customfigure2}{HTML}{FFD6AA} 

\title{Foundation Models for Recommender Systems: A Survey and New Perspectives}

\author{
Chengkai Huang$^1$
\and
Tong Yu$^2$\and
Kaige Xie$^{3}$\and
Shuai Zhang$^4$ \and
Lina Yao$^{1,5}$ \And
Julian McAuley$^6$
\affiliations
$^1$The University of New South Wales \qquad
$^2$Adobe Research \\
$^3$Georgia Institute of Technology \qquad
$^4$ETH Zurich \\
$^5$CSIRO’s Data61 \qquad
$^6$University of California San Diego
\emails
\{chengkai.huang1, lina.yao\}@unsw.edu.au,
tyu@adobe.com,
kaigexie@gatech.edu,
cheungshuai@outlook.com,
lina.yao@data61.csiro.au,
jmcauley@eng.ucsd.edu
}

\begin{document}

\maketitle

\begin{abstract}

Recently, Foundation Models (FMs), with their extensive knowledge bases and complex architectures, have offered unique opportunities within the realm of recommender systems (RSs). In this paper, we attempt to thoroughly examine FM-based recommendation systems ({\bf FM4RecSys}).
We start by reviewing the research background of FM4RecSys. Then, we provide a systematic taxonomy of existing FM4RecSys research works, which can be divided into four different parts including data characteristics, representation learning, model type, and downstream tasks. Within each part, we review the key recent research developments, outlining the representative models and discussing their characteristics. Moreover, we elaborate on the open problems and opportunities of FM4RecSys aiming to shed light on future research directions in this area. In conclusion, we recap our findings and discuss the emerging trends in this field.
   
\end{abstract}

\section{Introduction}\label{sec:intro}

Recommender Systems (RSs) tailor content and experiences to personalized preferences, increasingly contributing to the business enhancement and decision-making processes~\cite{RicciRS15,ZhangYST19}.
In parallel, foundation models (FMs) have made significant strides in areas such as natural language processing, computer vision, and multi-modal tasks. Recently, FMs have been reshaping recommender system architectures, enhancing performance, and offering new interaction methods in RecSys. Foundation model-based recommender systems with enhanced generalization capabilities are apt at leveraging more complex user-item information and handling more diverse RS tasks~\cite{geng2022recommendation}. To be specific, Foundation Models for Recommender Systems (\textbf{FM4RecSys})
refer to leveraging the knowledge from pre-training and recommendation datasets to capture rich representations of user preferences, item features, and contextual variables for improving personalization and prediction accuracy in recommendation tasks. Next, we will explore the motivations of existing works to deepen the understanding of the application and impact of FMs in this context.

\subsection{Motivation}

We enumerate the primary motivations driving the research in the evolving landscape of FM4RecSys.
\paragraph{Enhanced Generalization Capabilities.} Foundation Models are designed to learn from large-scale data, enabling them to understand complex patterns. FMs can generalize better to new, unseen data~\cite{foudnation_survey8}. In the context of RSs, this means that FMs can more accurately predict user preferences and behaviors, especially in scenarios with sparse data or novel items (defined as zero-shot/few-shot recommendations in some papers~\cite{gao2023chat,zeroshotRec,hou2023large}), 
By inferring user preferences or item characteristics from limited information or interactions, the recommendations can be made more effective even for new users or items.

\begin{figure*}[h]
    \centering
    \includegraphics[width=1\textwidth]{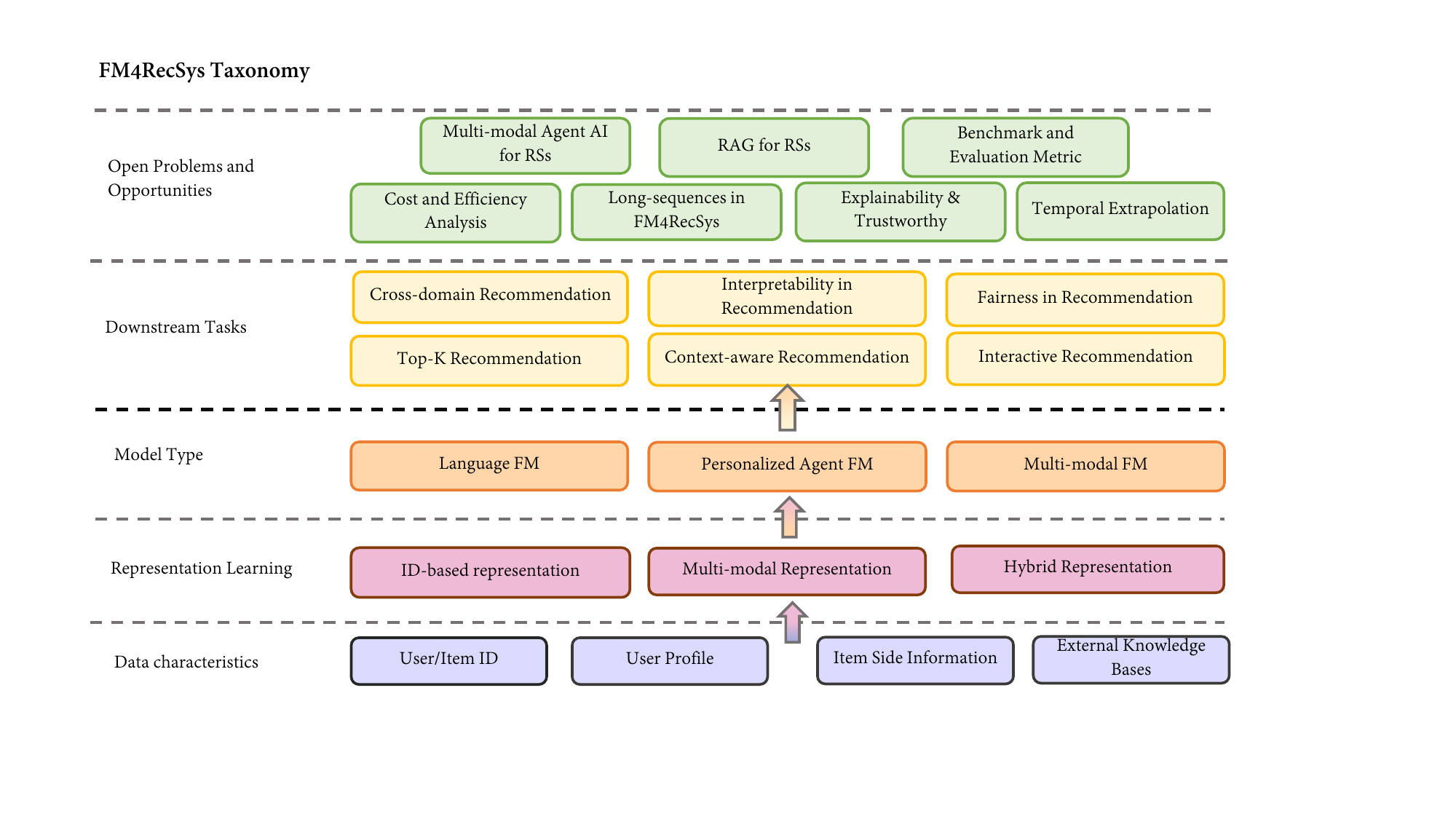} 
    \caption{The taxonomy of FM4RecSys from data characteristics to open problems and opportunities. 
    }    
    \label{fig:overview}
\end{figure*}

\paragraph{Elevated Recommendation Experience} Foundation Models inaugurate a transformative interface paradigm for recommendation systems, significantly altering the user interaction experience. 
For instance, conversational RS is a classic use case, previous CRSs~\cite{GaoLHRC21,Lei0MWHKC20} 
predominantly rely on pre-established dialogue templates, a dependency that often constrains the breadth and adaptability of user engagements. 
In contrast, FMs introduce a paradigm shift towards more dynamic and unstructured conversational interactions, offering enhanced interactivity and flexibility.
The interactive design allows for more engaging and natural user interactions with the system. Users can conversationally communicate their preferences, ask questions, and receive customized recommendations.

\paragraph{Improved Explanation \& Reasoning Capabilities.} Foundation Models augment the explanation and reasoning capabilities. Whereas traditional recommender systems predominantly derive explanations from rudimentary sources such as user reviews or elementary user behaviors, including co-purchased items or peer purchases, these explanations are often bereft of in-depth logic and context~\cite{LiZC20}. 
In contrast, Foundation Models possess the ability to formulate explanations that are enriched with a comprehensive grasp of commonsense and user-specific context~\cite{reasoning_survey}. These models leverage an array of data, encompassing user preferences, historical interactions, and distinctive item characteristics, to generate explanations that are both more coherent and logically sound.
Utilizing Foundation Models to deeply interpret user behavior sequences and interests can significantly enhance the effectiveness of future recommender systems in complex scenarios~\cite{Yanreasoning}. This methodology promises to advance informed and responsible decision-making processes in areas like medicine and healthcare, e.g., treatment and diagnosis recommendations.

\subsection{Distinguishing Features from Recent LLM based Recsys Surveys}

Multiple reviews have been conducted on the intersection of LLMs and RecSys. 
Liu \textit{et al.}~\shortcite{survey2} delves into the training strategies and learning objectives of language modeling paradigm adaptations for recommenders, while Wu \textit{et al.}~\shortcite{survey3} provides insights from both discriminative and generative viewpoints on Language Model-based Recommender Systems (LLM4Rec). Lin \textit{et al.}~\shortcite{survey1} introduces two orthogonal perspectives: where and how to adapt LLMs in recommender systems. Fan \textit{et al.}~\shortcite{survey4} offers an overview of LLMs for recommender systems, concentrating on paradigms such as pre-training, fine-tuning, and prompting. Lin \textit{et al.}~\shortcite{survey5} summarizes the current progress in generative recommendations, organizing them across various recommendation tasks.

\paragraph{Differences and Key Contributions:}In contrast to previous surveys, our methodology introduces a unique viewpoint for examining the intersection of FM4RecSys. As shown in Figure~\ref{fig:overview}, we systematically outline the framework for using Foundation Models (FMs) in recommender systems (FM4RecSys), covering everything from the characteristics of recommendation data to specific downstream tasks. Our approach to categorizing FM4RecSys is two-pronged, focusing on both the types of models used and the recommendation tasks themselves. This survey goes beyond previous surveys by not only covering large language models (LLMs) but also including a wider array of foundation models. We further delve into the latest unresolved questions and potential opportunities in this area. 

\section{Research Progress in FM4RecSys}

\subsection{Data Characteristics and Representation Learning}
\label{sec:unidata}

In the pre-foundation model era, recommender systems heavily relied on user and item representation from one-hot encoding. With the advent of FM4RecSys,
there is a shift towards embracing more diverse inputs such as user profiles, item side information, and external knowledge bases like WIKIPEDIA for enhanced recommendation performance. 
To be specific, numerous works
~\cite{BaoZZWF023,HuaLXCZ23} have identified that the key to building FM-based recommenders lies in bridging the gap between FMs' pre-training and recommendation tasks. To narrow the gap, existing work usually represents recommendation data in natural language for fine-tuning on FMs~\cite{Yaochen}. 
In this process, each user/item is represented by a unique identifier (e.g., user profile, item title, or numeric ID), and subsequently, the user’s historical interactions are converted into a sequence of identifiers. FMs can be fine-tuned on these identifiers to learn their representations to excel at recommendation tasks. 
Current recommendation data representation methods can be categorized as ID-based representation, multi-modal representation, and hybrid representation.

In FM context, recent studies on ID-based representation utilize numeric IDs like ``[prefix]+[ID]'' (e.g., ``user\_123'' or ``item\_57'') to represent users and items, effectively capturing the uniqueness of items~\cite{geng2022recommendation,hua2023index}. Nevertheless, numeric IDs lack semantics and fail to leverage the rich knowledge in FMs. Furthermore, FMs require sufficient interactions to fine-tune each ID representation, limiting their generalization ability to large-scale, cold-start, and cross-domain recommendations. 
Additionally, ID indexing necessitates updates to vocabularies to handle out-of-vocabulary (OOV) issues and parameter updates of the FMs that incur extra computational costs, highlighting the need for more informative representations.

A promising alternative way lies in leveraging multi-modal side information, including utilizing images~\cite{SarkarBVLBLM23} (such as item visuals), textual content~\cite{recformer,ZhangW23} 
(encompassing item titles, descriptions, and reviews), multi-modal elements~\cite{ShenZXJ22,YouwangKO22} 
(like short video clips and music), and external knowledge sources~\cite{ZhaiZWL023,Yunjia} (such as item relationships detailed in Wikipedia). 
Yuan \textit{et al.}~\shortcite{YuanYSLFYPN23} underscores the advantages of multi-modality-based RS when compared to ID-based counterparts, drawing attention to the performance gains.

However, the alignment between pure item side information and user-item interactions may not always be consistent~\cite{Yaochen,Jiayi}. In other words, two items with similar visual or textual features might not necessarily share similar interaction patterns with users.
Thus, another way to utilize the hybrid representation is to combine the ID and multi-modal side information to achieve both distinctiveness and semantic richness. For instance, TransRec~\cite{Xinyu} utilizes multi-faceted identifiers that combine IDs, titles, and attributes to achieve both uniqueness and semantic richness in item representation. CLLM4Rec~\cite{Yaochen} extends the vocabulary of FMs by incorporating user/item ID tokens and aligning them with user-item review text information through hard and soft prompting, allowing for accurate modeling of user/item collaborative information and content semantics. 

\subsection{Classification Framework for FM4RecSys}\label{sec:tax}

\begin{figure}[!h]  
	\centering
        \includegraphics[width=1\linewidth]{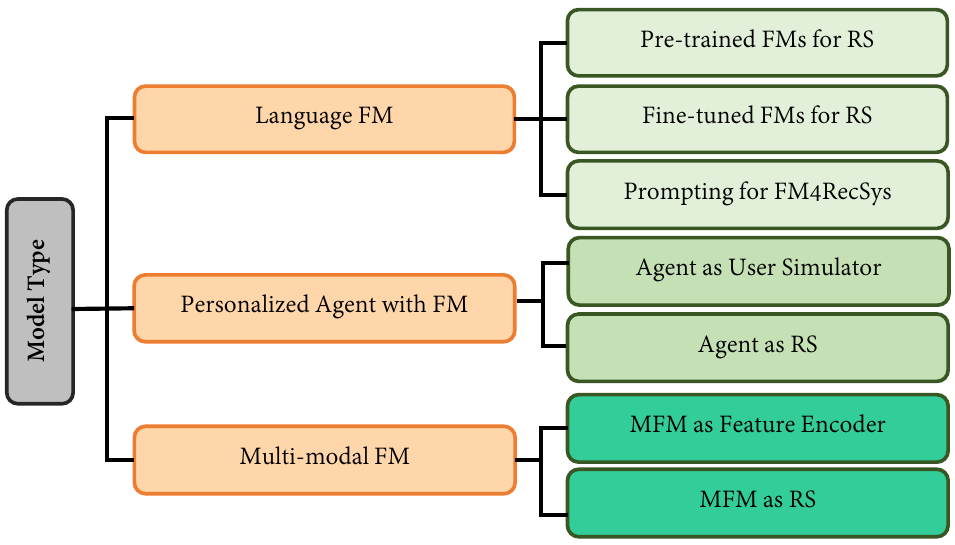}
	\caption{Classification frameworks in FM4RecSys from the Model Type perspective. 
 }
\label{fig:overview2}
\end{figure}

Our structured overview of the FM4RecSys classification frameworks, as presented in Figure~\ref{fig:overview2}, is organized by model type. Unlike earlier surveys focused on LLMs, our framework expands to include discussions on the latest Language Foundation Models for RSs, as well as additional research on multi-modal FM-based systems and personalized agents using FMs in RSs. This broader inclusion allows for a more comprehensive understanding of the current state and potential advancements in the field of FM-based RSs.

\subsubsection{Language Foundation Models for RecSys}\label{sec:language}

Language Foundation Models are the main branch in FM4RecSys, focusing on {\it pre-trained} and {\it directly fine-tuned models}, and {\it prompting techniques}.

\par {\it Pre-trained models for FM4RecSys.} Few works like M6 Rec~\cite{cui2022m6} and PTUM~\cite{wu2020ptum} pre-train the whole model on massive recommendation datasets by adopting transformer-based models for next-item prediction and applying different language modeling tasks, such as masked language modeling, permutation language modeling and so on. This line of work generally requires a large amount of domain data for recommender systems, leading to high training costs.

\par {\it Directly fine-tuned models for FM4RecSys.} A series of works adopt the fine-tuned FMs as RS. 
InstructRec~\cite{InstruRec} designs abundant instructions for tuning, including 39 manually designed templates with preference, intention, task form, and context of a user. 
After instruction tuning, LLMs can understand and follow different instructions for recommendations.
TallRec~\cite{TallRec} uses LoRA~\cite{lora}, a parameter-efficient tuning method, to handle the two-stage tuning for LLMs. It is first fine-tuned on general data of Alpaca, and then further fine-tuned with the historical information of users. It utilizes item titles as the input and shows effectiveness for cold-start recommendations.
BIGRec~\cite{BigRec} emphasizes that LLMs struggle to integrate statistical data such as popularity and collaborative filtering because of their inherent semantic biases. To address this, BIGRec fine-tunes LLMs through instruction tuning to produce tokens that symbolize items. However, aligning LLM outputs with real-world items is challenging due to their inventive nature. BIGRec subsequently aligns these generated tokens with real items in the recommendation database by incorporating statistical data like item popularity. 

\par {\it Prompting for FM4RecSys.} Another approach involves non-tuning paradigms, where LLM parameters remain unchanged, and the focus is on extracting knowledge using prompting strategies. 
Existing work of non-tuning paradigm focuses on designing appropriate prompts to stimulate the recommendation abilities of LLMs.
Liu \textit{et al.}~\shortcite{benchmarking} proposes a prompt construction framework to evaluate the ability of ChatGPT on five common recommendation tasks, providing zero-shot and few-shot versions for each type of prompt. He \textit{et al.}~\shortcite{hou2023large} not only uses prompts to evaluate the ability of LLMs on sequential recommendation but also introduces recency-focused prompting and in-context learning strategies to alleviate order perception and position bias issues of LLMs.
More recently, some works~\cite{prompting_survey} have focused on designing novel structures in prompting for FM4RecSys. 
Yao \textit{et al.}~\shortcite{knowledgeplugin} includes heuristic prompts including item attributes in natural language, along with collaborative filtering information presented through text templates and knowledge graph reasoning paths. Similarly, Rahdari \textit{et al.}~\shortcite{Behnam23} has crafted hierarchical prompt structures that encapsulate information about recommended items and top-k similar item information in the user interaction history.

\subsubsection{Personalized Agent for RecSys with Foundation Models}\label{sec:agent}

\begin{figure*}[!h]  
	\centering
        \includegraphics[width=1\linewidth]{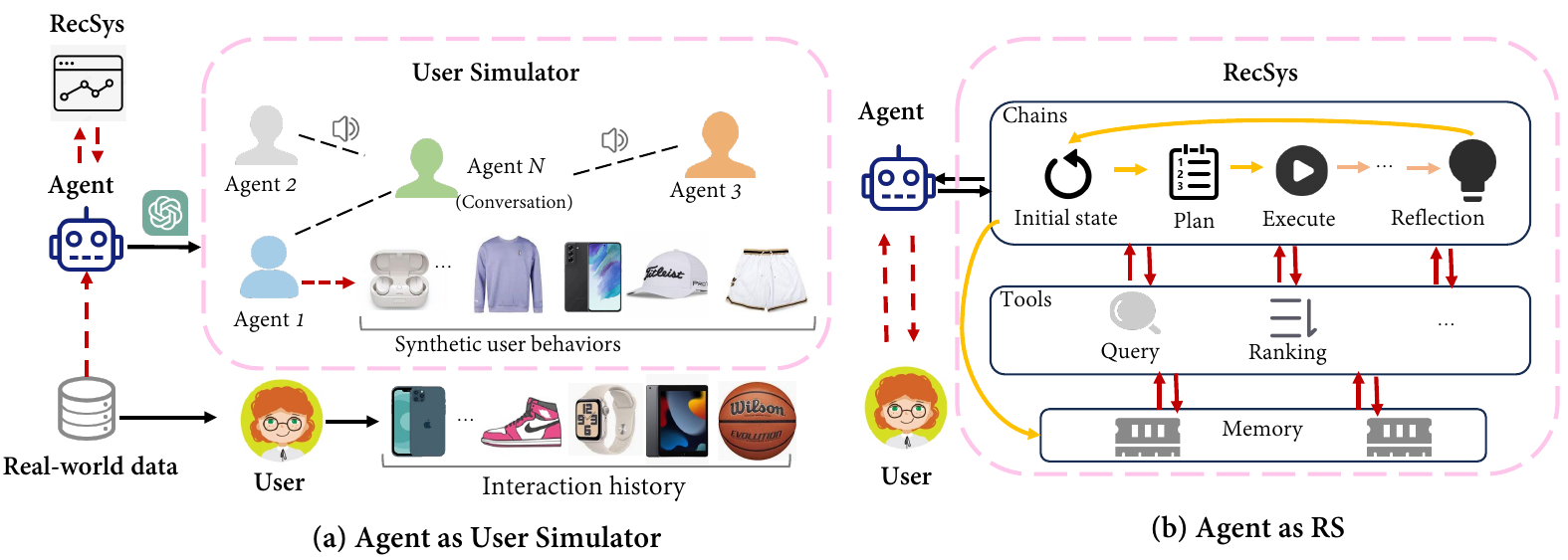}
	\caption{Two types of personalized agents in FM4RecSys: (a) Agent as User Simulator and (b) Agent as Recommender System. }
	\label{fig:agent}
\end{figure*}

Agents are typically represented as either User Simulators or the Recommender System itself as shown in Figure~\ref{fig:agent}. 

\par {\it {Agent as User Simulator}.} 
using agents to simulate user behaviors for real-world recommendation.
Gathering sufficient and high-quality user behavior data is expensive and ethically complex. Besides, traditional methods~\cite{ZhuLLWGLC17,RecSim} 
often struggle to simulate complex user behaviors while FMs show potential in simulating user behaviors~\cite{wang2023recagent}. Consequently, employing personalized agents powered by FMs for RSs emerges as a logical and effective strategy.
Wang \textit{et al.}~\shortcite{wang2023recagent} treats each user as an FM-based autonomous agent within a virtual simulator named RecAgent. This simulator allows for the free interaction, behavior, and evolution of different agents, taking into account not only actions within the RS like item browsing and clicking but also external factors like social interactions. Zhang \textit{et al.}~\shortcite{agent4rec} further investigates the extent to which FM-empowered generative agents can accurately simulate real human behavior within the movie RS. They design Agent4Rec, a recommender system simulator with 1,000 LLM-empowered generative agents interacting with personalized movie recommendations in a page-by-page manner with various actions. 
After that, \cite{AgentCF} presents simulating user-item interactions in RS, treating both users and items as agents and enabling a collaborative learning process that optimizes the interactions between agents.

\par {\it {Agent as RecSys}.} leveraging the robust capability of FMs, including reasoning, reflection, and tool usage for recommendation.
Wang \textit{et al.}~\shortcite{wang2023recmind} first introduces a Self-Inspiring planning algorithm that keeps track of all past steps of the agent to help generate new states. At each step, the agent looks back at all the paths it has taken before to figure out what to do next. This approach aids in employing databases, search engines, and summarization tools, combined with user data, for producing tailored recommendations.
\cite{InteRecAgent} model the FMs as the brain, while recommendation models serve as tools that supply domain-specific knowledge, then FMs can parse user intent and generate responses.
They specify a core set of tools essential for RS tasks—Information Query, Item Retrieval, and Item Ranking—and introduce a candidate memory bus, allowing previous tools to access and modify the pool of item candidates.

\subsubsection{Multi-modal Foundation Models for RecSys}\label{sec:multi-modal}
 
The use of Multi-modal Foundation Models (MFMs) in RecSys falls into two primary categories: employing MFMs as encoders for multi-modal features, and integrating MFMs centrally in recommender systems for direct multi-modal data processing and recommendation generation.

\par {\it MFM as Feature Encoder.} MFMs serve as encoders for multi-modal data, leveraging their robust representation and generalization capability to extract features for recommendation. Examples include ViT~\cite{JiLZWNW23} for image features of items, CLIP~\cite{WangZWWLLYZZX23,ZhangSGLLL023} 
for both image and text features, SentenceBert/BLOOM-176B~\cite{ShenZXJ22} for text, DeepSim~\cite{McKeeSSR23} for audio, and SlowFast~\cite{ShenZXJ22} for video features. 
A key challenge for these methods is how to bridge the disparity between the optimization goals of pre-training and those of downstream recommendation tasks.

\par {\it MFM as RecSys.} Geng \textit{et al.}~\shortcite{geng2023vip5} introduced VIP5, an extension of their earlier work, focusing on both visual and textual modalities in recommendation tasks. VIP5 incorporates multi-modal personalized prompts and a parameter-efficient training strategy, which involves freezing the foundational P5 backbone and fine-tuning lightweight adapters for improved performance and efficiency. Building on the same foundational work, 
Zhai \textit{et al.}~\shortcite{ZhaiZWL023} presents the KP4SR approach, which utilizes an external knowledge base and structured knowledge prompts to address the semantic gap in sequential recommendation systems. In a recent exploration, Zhou \textit{et al.}~\shortcite{PeilinGPT4V} investigates GPT-4V's application in visually-informed recommendation tasks, assessing its zero-shot recommendation capability across various domains, including culture, art, entertainment, and retail. However, as an initial study, it faces limitations such as the absence of quantitative assessment, sample bias, and potential response inconsistencies.
 
\subsection{Applications of FM4RecSys}\label{sec:application}

\textbf{Top-K Recommendation Task}

\noindent  The top-K recommendation task is essentially ranking. However, if user information (including meta-information and item interaction history) is overly lengthy, it may exceed the input length capacity of foundation models. To address this, methods based on ID representation can be employed~\cite{hua2023index} in FMs. Foundation models use a prompt that only includes user information, asking the foundation models to directly generate recommendations for those users~\cite{Xu2023,geng2022recommendation}.
In the case of multi-modal and generative representation methods, the generated recommendation items can undergo similarity calculations with the multi-modal representation of ranking candidates~\cite{LiuMXLYL0023}. Additionally, some approaches~\cite{LiZM23,DaiSZYSXS0X23} follow practices from the NLP field. They select K-item negative samples or hard examples, feed them along with user prompts to FMs, and obtain the final rankings. However, these approaches target the idealized experimental scenario and may not be practical for real-world recommendation systems with millions of items. 

\subsubsection{FM for Context-aware RecSys}\label{sec:context}

Various FM-based approaches have been proposed to exploit their capability in the realm of context-aware recommendation. Not only can the extensive world knowledge stored in FMs serve as a rich source of background information for items~\cite{HarteZLKJF23}, but also can the reasoning capability of FMs augment the next item prediction~\cite{Yunjia,Wangyureasoning}. 
\cite{HarteZLKJF23} first explore three different methods of utilizing foundation models' knowledge for context-aware recommendations, based on FM semantic similarity, FM prompt fine-tuning, and BERT4Rec initialized by FM semantic embedding.
Wu \textit{et al.}~\shortcite{YiqingPPR} generate personalized soft prompts using user profile knowledge and employ prompt-oriented contrastive learning for effective training.
After that, Zhai \textit{et al.}~\shortcite{ZhaiZWL023} introduce knowledge prompt-tuning for context-aware recommendations, which effectively integrates external knowledge bases with FMs, transforming structured knowledge into prompts to refine recommendations by bridging semantic gaps and reducing noise.  
More recently, Liao \textit{et al.}~\shortcite{Jiayi} employ a hybrid approach for item representation in input prompts for FM, combining ID-based item embeddings from traditional recommenders with textual item features, bridging the modality gap between traditional recommender systems and FMs through an adapter, and facilitating the transfer of user behavioral knowledge to the FM's input space.
Meanwhile, Wang \textit{et al.}~\shortcite{Wangyureasoning} utilize the reasoning capability of Foundation Models (FMs) and introduce a collaborative in-context demonstration retrieval method, abstracting high-level user preferences and reducing noise to improve the recommendation process without the need for FM fine-tuning.

\subsubsection{FM for Interactive RecSys}\label{sec:interactive}

The goal of interactive recommendation is to not only suggest items to users over multiple rounds of interactions, but also provide human-like responses for multiple purposes such as preference refinement, knowledgable discussion, or recommendation justification~\cite{JannachMCC21,SunZ18}. 
The emergence of FMs has undoubtedly impacted interactive RS, especially CRS-related research.
He \textit{et al.}~\shortcite{HeXJSLFMKM23} presents empirical evidence that FMs, even without fine-tuning, can surpass existing conversational recommendation models in a zero-shot setting.
After that, a series of works~\cite{liu2023chatgpt,spark,wang2023recmind,KyleCRS} 
adopt the role-playing prompt to guide ChatGPT/GPT-4 in simulating user interaction with conversational recommendation agents. These works augment FMs' capability through techniques such as RAG and Chain-of-Thought (CoT).
Meanwhile, several studies are built based on prior work~\cite{ZhouZBZWY20} in knowledge graph-based interactive recommendation. For instance, Wang \textit{et al.}~\shortcite{lingzhi2021} introduce a framework that integrates PLMs like DialoGPT with a knowledge graph to generate dialogues and recommend items, showcasing how FM's generative capability can be utilized for interactive recommendation.
Zhang \textit{et al.}~\shortcite{Gangyi} explore a user-centric approach, emphasizing the adaptation of FMs to users' evolving preferences through graph-based reasoning and reinforcement learning. Recently, 
Wang \textit{et al.}~\shortcite{wang2023rethinking} critique the current evaluation protocols for interactive RSs and introduce an FM-based user simulator approach, iEvaLM, which significantly enhances evaluation accuracy and explainability. However, FMs for interactive recommendation are still hindered by a tendency towards popularity bias and sensitivity to geographical regions. 

\subsubsection{Cross-domain in FM4RecSys}

In real-world scenarios, data sparsity is a pervasive issue for Collaborative Filtering (CF) recommender systems, as users rarely rate or review a broad range of items, particularly new ones. Cross-domain recommendation (CDR) tackles this by harnessing abundant data from a well-informed source domain to enhance recommendations in a data-scarce target domain. Multi-domain recommendation (MDR) extends this concept by utilizing auxiliary information across multiple domains to recommend items within those domains to specific users~\cite{ZhuW00L021}. 
However, domain conflicts remain a significant hurdle, potentially limiting the effectiveness of recommendations. The advent of foundation models that are pre-trained on extensive data across various domains and possess the cross-domain analogical reasoning ability~\cite{HuSLC23} presents a promising solution to these challenges.

HAMUR~\cite{Xiaopeng} designs a domain-specific adapter to be integrated into existing models and a domain-shared hyper-network that dynamically generates adapter parameters to tackle the mutual interference and the lack of adaptability in previous models.
Tang \textit{et al.}~\shortcite{Zuoli} discuss the application of FMs in multi-domain recommendation systems by mixing the user behavior across different domains, concatenating the title information of these items into a sentence, and modeling the user behavior with a pre-trained language model, which demonstrates the effectiveness across diverse datasets. 
The S\&R Multi-Domain FM~\cite{GongDSSLZ23} employs FMs to refine text features from queries and items, improving CTR predictions in new user or item scenarios. 
KAR~\cite{Yunjia} further leverages the power of FMs for open-world reasoning and factual knowledge extraction and adaptation. It introduces a comprehensive three-stage process encompassing knowledge reasoning and generation, adaptation, and subsequent utilization.
Based on the S\&R Multi-Domain FM, Uni-CTR~\cite{Zichuan} employs a unique prompting strategy to convert features into a prompt sequence that FMs can use to generate semantic representations, capturing commonalities between domains while also learning domain-specific characteristics through domain-specific networks.  
More recently, Fu \textit{et al.}~\shortcite{Junchen} investigate the efficacy of adapter-based learning for CDR, which is designed to leverage raw item modality features, like texts and images, for making recommendations. They conduct empirical studies to benchmark existing adapters and examine key factors affecting their performance.

\subsubsection{Interpretability and Fairness in FM4RecSys}\label{sec:fairness}

\par {\it Interpretability in FM4RecSys.} A common task in enhancing the interpretability of recommendation systems is the generation of natural language explanations~\cite{ZhangC20}. This involves directing the recommender or external model to produce, in a sentence or a paragraph, the reasons behind recommending a specific item to a particular user. For instance, given a user $u$ and an item $i$, the model is tasked to generate a coherent and understandable explanation in natural language that elucidates why item $i$ is recommended to user $u$. 
A series of works use ID-based representation and leverage prompts like ``explain to the user $u$ why item $i$ is recommended''~\cite{LiZC20}.
However, the use of IDs alone in prompts may sometimes lead to vague explanations, lacking clarity on specific aspects of the recommendation. To address this, Cui \textit{et al.}~\shortcite{cui2022m6} propose to integrate item features as hint words in the prompt, aiming to guide the model more effectively in its explanatory process.
Recently,  Liu \textit{et al.}~\shortcite{liu2023first} leverage continuous prompt vectors instead of discrete prompt templates.
Remarkably, it is found that ChatGPT, operating under in-context learning without fine-tuning, outperforms several traditional supervised methods~\cite{liu2023chatgpt}. 

\par{\it Fairness in FM4RecSys.} The imperative of fairness in RS stems from its widespread use in decision-making and meeting user demands. Nonetheless, there currently exists a deficiency in comprehending the degree of fairness manifested by foundation models in recommendation systems, as well as in identifying suitable methodologies for impartially addressing the needs of diverse user and item groups within these models~\cite{abs-2305-12090,ZhangBZWF023}.
For the user group side, Hua \textit{et al.}~\shortcite{abs-2305-12090} propose the Unbiased Foundation Model for Fairness-aware Recommendation (UP5) based on Counterfactually-Fair-Prompting (CFP) techniques. After that, Zhang \textit{et al.}~\shortcite{ZhangBZWF023} craft metrics and a dataset that accounts for different sensitive attributes in two recommendation scenarios: music and movies, and evaluate the ChatGPT's fairness in RS concerning various sensitive attributes on the user side. 
For the item side, Hou \textit{et al.}~\shortcite{hou2023large} guide FMs with prompts to formalize the recommendation task as a conditional ranking task to improve item-side fairness. Research on non-discrimination and fairness in FM4RecSys is still in its early stages, indicating a need for further investigation.

\section{Open Problems and Opportunities }
\subsection{Long-sequences in FM4RecSys}

FM4RecSys faces challenges with lengthy input sequences due to their fixed context window limitations, impacting their efficacy in tasks requiring extensive context~\cite{kitaev2019reformer,beltagy2020longformer}, like in context-aware recommendations.
Context-aware RSs, relying on a user's comprehensive interaction history and extensive item ranking lists, often exceed the FMs' context capacity, leading to less effective recommendations. Adaptations from NLP techniques are being explored, including segmenting and summarizing inputs to fit within the context window and employing strategies like attention mechanism and memory augmentation to enhance the focus on pertinent segments of the input. The RoPE technique~\cite{su2024roformer}, with its innovative rotary position embedding, shows promise in managing long inputs and offers a potential solution for maintaining RSs' performance despite the context window constraint of FMs.

\subsection{Explainability and Trustworthyness}
Enhancing explainability and trustworthiness in RS is always a significant challenge, especially in the FM era. The complexity and size of FMs introduce new hurdles in explaining FM4RecSys.
There are two primary approaches to explainability in RS: one involves generating natural language explanations for recommendations, and the other dives into the model's internal workings. 
The former approach has seen considerable explorations~\cite{zhang2020explainable} in the pre-FM era whereas the latter is less developed.
There are also some works~\cite{Behnam23,Yanreasoning}
that align FMs such as their prompts, with explicit knowledge bases like knowledge graphs. This alignment can make the model’s decision-making process traceable as specific paths in the knowledge graph, offering a clearer explanation. 
However, these approaches are still in their preliminary phase and might be further enhanced by techniques such as Chain/Tree of Thoughts. 

\subsection{Temporal Extrapolation}

Recent studies~\cite{DBLP:journals/corr/abs-2310-10196} have demonstrated that FMs can extrapolate time series data in a zero-shot manner, achieving performance on par with or superior to specialized models trained on specific tasks. This success is largely due to FM's ability to capture multi-modal distributions and their inclination towards simplicity and repetition, which resonates with the repetitive and seasonal trends commonly observed in time series data. Time series modeling, distinct from other sequence modeling due to its variable scales, sampling rates, and occasional data gaps, has not fully benefited from large-scale pretraining. To address this, LLMTIME2~\cite{gruver2023large} leverages LLMs for continuous time series prediction by encoding time series as numerical strings and treating forecasting as a next-token prediction task. This method, which transforms token distributions into continuous densities, facilitates an easy application of LLMs to time series forecasting without the need for specialized knowledge or high computational costs, making it particularly beneficial for resource-constrained scenarios. Moreover, by viewing user preference data as a time series sequences, these models might adeptly adapt to long-term shifts in preferences and enhance personalization and predictive accuracy over time, especially with the zero-shot capability of approaches like LLMTIME2, which enables a rapid adaptation to user preference change without the need for extensive retraining.

\subsection{Multi-modal Agent AI for RecSys}

Multi-modal Agent AI~\cite{durante2024agent} is an emerging field that focuses on AI systems that can perceive and act in various domains and applications. Aiming to achieve intelligence based on a multi-modal understanding of the surrounding world, agent AI systems leverage a variety of generative models and data sources for reality-agnostic training. These systems can be embodied in physical and virtual environments, allowing them to process visual and contextual data, understand user actions and behaviors, and produce meaningful responses. 
In the application of RSs, an agent can make decisions on what to recommend based on inferences about user preferences. An agent can also be made more interactive to leverage real-time responses or feedback from users or environments to adjust inferences and improve recommendations. 
In particular, not only can they serve as simulators for RS, but also act as simulators for users. 
This approach allows for data collection and training in an offline environment, reducing the A/B testing  costs in the real world. It can be extended to a broader range of users, for example, in applications like route planning recommendations and medical drug discovery and recommendations.

\subsection{RAG meets RecSys}

Retrieval-Augmented Generation (RAG) is a technique used in FMs to enhance their generative capability by integrating external data retrieval into the generative process~\cite{ragsurvey}.
This approach improves the accuracy, credibility, and relevance of FM outputs, particularly in knowledge-intensive tasks like information retrieval and RS. RAG aims to address outdated knowledge, the generation of incorrect information (hallucinations), and limited domain expertise by combining the FM's internal knowledge with dynamic external knowledge bases.
RAG is suitable for enhancing the FM4RecSys, especially in modeling lifelong user behavior sequences in real-world RS environments~\cite{ragpaper}. 
It could potentially ensure that the RecSys remains up-to-date with continuous shifts in user preferences and trends, which is critical for precise identification and documentation of long-term behavioral patterns. For instance, considering the input token length restriction of FMs, RAG may be utilized to selectively extract pertinent portions of a user's interaction history and associated external knowledge, thereby conforming to the model's input constraint. Additionally, RAG may lessen the likelihood of producing irrelevant recommendations or non-existent items (hallucinations), thereby enhancing the reliability of FM4RecSys.

\subsection{System Performance Analysis: APIs cost, Training \& Inference Efficiency}
In the development of FM-based RSs, a critical aspect is the cost assessment, which varies depending on the data and model selections throughout the training and inference phases~\cite{foudnation_survey8}. The training phase incurs costs due to the recommendation model's pretraining, fine-tuning, and algorithmic development, where the complexity and the need for specialized engineering can drive up expenses. During the recommendation inference stage, costs persist in the form of system upkeep, updates, and computational demands of API-driven service provision. For instance, systems like OpenAI's GPT-3/4~\cite{brown2020language,GPT4report} have costs associated with API usage and token interactions, escalating with more intricate or extensive usage. 
Furthermore, the incorporation of RAG tools can further elevate expenses by extending prompt lengths and, consequently, the number of tokens processed, leading to higher API fees. Additionally, customization through fine-tuning also adds to the overall expenses. 

\begin{table*}[!h]
\centering
\small 
\begin{tabular}{ccc} 
\toprule
\textbf{Cost and Efficiency} & \textbf{Possible Methods} & \textbf{References} \\ \midrule

\multirow{3}{*}{Training Cost} 
&  Data Selection for Pre-training &  \cite{GlassGCFPBGS20,DataSelection2} \\
&  Data Selection for Fine-tuning &  \cite{Chunting,Yihan} \\
                               & Parameter-efficient Fine-tuning    & \cite{Junchen} \\
\midrule                              
\multirow{2}{*}{Inference Latency} & Embedding Caching & \cite{hou2023learning,HarteZLKJF23} \\
                               & Lightweight FMs   &   \cite{ChenFC0ZWC20}, \cite{Xiaoqi}, \cite{Jilin} \\
\midrule                            
\multirow{3}{*}{API Cost} & Data Selection   &  \cite{HaoChen} \\
                               & Prompt Selection    &\cite{Jessewu} \\  & Adaptive RAG     &\cite{Mallen} \\              
\bottomrule
\end{tabular}
\caption{An organization of representative methods for reducing the cost and improving the efficiency for FM4RecSys.}
\label{cost_table}
\end{table*}


                               
                               
                               

Addressing efficiency in FM4RecSys is a practical challenge with direct implications for system performance and resource utilization. Referencing Table~\ref{cost_table}, we outline targeted solutions:

{\it 1) Training Cost Reduction:} For pre-training or fine-tuning Foundation Models within recommender systems, it is necessary to carefully select the most informative and diverse data so that the model can efficiently capture essential user-item interaction patterns and features and accelerate the learning process~\cite{GlassGCFPBGS20,DataSelection2}.
Additionally, employing techniques~\cite{Junchen} like LoRA~\cite{lora} and LoftQ~\cite{Yixiao} for fine-tuning can help in managing memory usage and reducing training time. 

{\it 2) Inference Latency Reduction:} The computational demand for FM inference is notable. Strategies such as employing pre-computed embedding 
caches~\cite{hou2023learning,HarteZLKJF23} (e.g., VQ-Rec or LLM4Seq) can offer some relief by speeding up inference. Similarly, efforts to compact the model size through 
distillation~\cite{Xiaoqi}, pruning~\cite{ChenFC0ZWC20}, 
and quantization~\cite{Jilin} can lead to improvements in memory cost and inference speed.

{\it 3) API Cost Reduction:} In FM-based API recommender systems, efficient data selection can enhance the fine-tuning efficiency by using a selected set of data points~\cite{HaoChen}. Additionally, refining prompt engineering with methods like prompt generation or compression~\cite{Jessewu} may lead to more efficient processing of FM inputs by making prompts more concise or better tailored, though the gains should be considered within realistic expectations. 
Besides, utilizing RAG to enhance API-based RSs can result in an additional context length, especially when retrieving longer item descriptions as prompt inputs. Therefore, adopting adaptive RAG~\cite{Mallen} is also an effective method to reduce API costs in that case.

\subsection{Benchmarks and Evaluation Metrics}

Liu \textit{et al.}~\shortcite{benchmarking} benchmark four state-of-the-art Large Language Models (LLMs) on five recommendation system tasks, using both quantitative and qualitative methods. However, they only focus on specific LLMs like ChatGPT and ChatGLM, limiting experiments to the Amazon Beauty dataset due to the high computational cost.
Thus, because of the domain-specific nature of recommendation systems, there is a need for more datasets, recommendation tasks, and evaluation metrics to create a more unified benchmark. 
Moreover, for multi-modal and personalized agents FMs, devising new benchmarks and evaluation metrics specifically for recommendation scenarios is essential.
In summary, to comprehensively evaluate and enhance the performance of FM-based recommendation systems, a holistic and diversified benchmark is essential. Such a benchmark should include a variety of datasets, diverse recommendation tasks, and metrics that are adaptable to different models. 

\subsection{Discussion on Emerging Trends}
We draw on a concise discussion of how the impressive understanding and generation power of FMs act as a dual-edged weapon in the context of FM4RecSys.

\par {\it From a perspective of safety}, FMs are vulnerable to red teaming attacks, where malicious actors craft poison prompts to manipulate the models into producing undesirable content. This content can range from fraudulent or racist material to misinformation or content inappropriate for younger audiences, potentially causing significant societal harm and putting users at risk~\cite{Boyi23}.
Thus, in the context of FM4RecSys especially when employing conversational interfaces, aligning FMs with human values becomes crucial. This alignment involves gathering relevant negative data and employing supervised fine-tuning techniques such as online and offline human preference training~\cite{YufeiAlign,Guohaihumanalign}. 
These methods can help in refining the models to adhere more closely to human instructions and expectations, ensuring the generative contents made by FM4RecSys are safe, reliable, and ethically sound.

\par {\it From a perspective of privacy,} if FMs are trained directly on a large amount of sensitive user interaction data, it might be possible for third parties to use methods like prompt injection to access specific user interaction histories, thereby constructing user profiles. In that sense, the incorporation of approaches such as federated learning~\cite{YangYu} and machine unlearning~\cite{Chen0ZD22} into FM4RecSys represents a promising direction for the future.

\section{Conclusion}
In this paper, we have furnished a thorough review of FM4RecSys, providing detailed comparisons and highlighting future research paths. 
We hope that this survey provides an overview of the challenges and the recent progress as well as some open questions and opportunities in foundation models to the RecSys research community.

\bibliographystyle{named}
\bibliography{long_ref}

\end{document}